\documentclass{aa501}
\usepackage{graphics}
\newcommand{\be}{\begin{equation}}
\newcommand{\ee}{\end{equation}}
\begin{document}

\title{Detection of a new, low-brightness supernova remnant possibly associated with EGRET sources}

\author{Jorge A. Combi$^{1}$, Gustavo E. Romero$^{1}$, Paula Benaglia$^{1}$ and Justin L. Jonas$^{2}$}
\offprints{J.A. Combi}

\institute{$^1$ Instituto Argentino de Radioastronom\'{\i}a,
C.C.5, (1894) Villa Elisa, Buenos Aires, Argentina \\
$^2$Department of Physics \& Electronics, Rhodes University,
Grahamstown 6140, South Africa}

\date{Received 19 September 2000 / Accepted 17 November 2000}

\titlerunning{Detection of a new supernova remnant}
\authorrunning{Combi et al.}

\abstract{ We report on the discovery of a shell-type supernova
remnant in the southern sky. It is a large ($\sim 8^{\circ} \times
8^{\circ}$), low-brightness source with a nonthermal radio
spectrum, which requires background filtering to isolate it from
the diffuse background emission of the Galaxy. Three 3EG
$\gamma$-ray sources are spatially correlated with the radio
structure. We have made 21-cm line observations of the region and
found that two of these sources are coincident with HI clouds. We
propose that the $\gamma$-ray emission is the result of hadronic
interactions between high-energy protons locally accelerated at
the remnant shock front and atomic nuclei in the ambient clouds.
}

\maketitle

\section{Introduction}

The origin of cosmic rays (CRs) is a long-standing problem in
contemporary astrophysics. It has been thought since the 1960s, on
the basis of energetic considerations, that protons are
accelerated up to energies of $\sim 10^{15}$ eV in supernova
remnants (SNRs) (Ginzburg \& Syrovatskii 1964). When first-order
Fermi acceleration at shock fronts was established as an efficient
mechanism for CR production (e.g. Bell 1978), SNRs became the most
natural candidates for the accelerating agents of the bulk of
galactic CRs. Multifrequency radio observations clearly show that
SNRs accommodate a population of relativistic electrons with the
expected energy distribution. However, there is at present not
such a conclusive observational evidence for shock acceleration of
protons and ions near SNRs.

High-energy $\gamma$-ray astronomy is the most appropriate tool to
probe SNRs in the search for an observable signature of proton
acceleration. The interactions between relativistic protons and
ambient nuclei can produce neutral pions ($\pi^0$) which decay to
yield $\gamma$-ray emission that might be detected by instruments
like the Energetic Gamma Ray Experiment Telescope (EGRET) on board
the Compton satellite. Possible associations of $\gamma$-ray
sources with SNRs have been claimed by several authors in the past
(Pollock 1985, Combi \& Romero 1995, Esposito et al. 1996, and
Combi et al. 1998a). However, a leptonic origin of the
$\gamma$-ray emission (e.g. Pohl 1996) cannot be ruled out in most
of these cases, especially when relatively young remnants are
considered.

In this paper we present results of a new observational study
designed to clarify the problem of the origin of galactic CRs.
Aharonian et al. (1994) and Aharonian \& Atoyan (1996) have
pointed out that the presence of nearby clouds can increase the
probability of detecting the nucleonic CR component locally
accelerated in a SNR because the clouds act as targets for the
relativistic protons, hence enhancing $\pi^0$ production and the
resultant $\gamma$-ray luminosity. A large remnant in a cloudy
medium may have several discrete $\gamma$-ray sources associated
with it. The study of the spectra and fluxes of these $\gamma$-ray
sources could provide information about the cosmic ray production
at several different sites where the clouds are overtaken by the
accelerated protons.

Some of the 170 unidentified 3EG $\gamma$-ray sources (Hartman et
al. 1999) appear to occur in small clusters subtending less than
$6^{\circ}$ in the sky at high galactic latitudes.  We have
searched the 408~MHz all-sky survey (Haslam et al. 1981) for radio
continuum emission towards the directions of those four clusters
of high-latitude sources that are visible from the southern
hemisphere. The EGRET confidence contours for the cluster
consisting of 3EG~J1834-2803, 3EG~J1847-3219, and 3EG~J1850-2652
correlate spatially with an extended shell-type radio source,
which is masked by steep gradients in the diffuse galactic
radiation. We have subsequently performed HI-line observations
towards this large area in the Capricornus region, and have
detected hydrogen clouds at the positions of two of the 3EG
sources. Additional radio continuum data was provided by the
Rhodes/HartRAO 2326~MHz survey (Jonas et al. 1998). Having radio
images at two widely-spaced frequencies allows the spatial
distribution of the spectral index to be computed, hence
establishing the nonthermal nature of the emission and revealing
the energy distribution of the electronic component of the CRs at
the SNR. In what follows we describe our data analysis procedures,
new observations, and discuss the results of our study.

\section{Data analysis and new observations}
\begin{figure}
\caption{Background filtered radio emission at 408 MHz of the
region surrounding the three $\gamma$-ray sources. Contours are
labelled in steps of 1 K in brightness temperature, starting at
1.2 K. The superposed grey-scaled levels represent the 99\%, 95\%,
68\%, and 50\% statistical probability that a $\gamma$-ray source
lies within each contour according to the EGRET catalog (Hartman
et al. 1999). HPBW $\sim 51^{'}$} \label{fig.1}
\end{figure}

The detection of weak, extended radio sources is a difficult task
due to the contamination produced by the diffuse background
radiation in the radio images. Faint sources with diameters of
several degrees can only be identified by using effective spatial
filtering techniques applied to good-quality, single-dish data. In
this study we have used data at 408-MHz and 2326-MHz from the
radio surveys by Haslam et al. (1981) and Jonas et al. (1998),
respectively. The angular resolutions (HPBW) of these surveys are
$51^\prime$ (408 MHz) and $20^\prime$ (2326 MHz). We have
extracted maps of $12^{\circ}\times12^{\circ}$ around the mean
position of the $\gamma$-ray sources and have then used the
background filtering method originally developed by Sofue \& Reich
(1979) in order to emphasize the fine radio structure superposed
on the diffuse galactic radio emission. This iterative procedure
results in the elimination of diffuse sources with angular scales
larger than a specified filtering beam (see Combi et al. 1998b for
more details). This technique is very versatile, can be used at
any galactic latitude, and is well-proven for a range of angular
scales. It was recently used to discover the radio counterpart of
the nearby SNR RX~J0852.0-4622, which is contaminated by the
bright, extended emission of the Vela region (Combi et al. 1999).

When the $12^{\circ}\times12^{\circ}$-maps are filtered, a steep
gradient of emission shows up towards the galactic plane. In order
to remove this residual background, we filtered a larger region of
$20^{\circ}\times24^{\circ}$, which includes the galactic plane.
Then the smaller maps were extracted again from the filtered,
larger map.

Figures 1 and 2 show the background filtered radio emission of the
region under study at 408- and 2326-MHz, respectively. A filtering
beam of \mbox{$240^\prime \times 240^\prime$} was applied to both
images (i.e. all emission structures with angular scales larger
than $4^\circ$ have been suppressed in these images). The
probability confidence contours of the 3EG sources have been
superposed in the maps. A large ($\sim 8^{\circ}\times8^{\circ}$),
shell-type source can be clearly seen at both radio frequencies.
In order to determine the spectral index distribution of the shell
emission we have convolved and re-tabulated the 2326-MHz data to
the same beam and tabular interval as the 408-MHz map. Following
Combi et al. (1998b), the spectral indices and the corresponding
errors were estimated from these maps. Uncertainties from the
filtering method also should contribute to the errors, but are
difficult to estimate. We think that a conservative assumption
here is to assume that the real errors double the formal ones. The
spectral index distribution, which is shown in Figure~3, is
consistent with independent estimates obtained using a different
filtering method developed by Jonas (1999).

\begin{figure}
\caption{Idem Fig. 1, but at 2.326 GHz. Brightness temperature
contours are in steps of 0.03 K, starting at 0.03 K. HPBW $\sim
20^{'}$} \label{fig.2}
\end{figure}

Aharonian \& Atoyan (1996) emphasize that the existence of a
potential CR accelerator, such as a SNR, is not a sufficient
condition for the generation of detectable levels of $\gamma$-ray
emission. Target nuclei are also required, so we have performed HI
observations towards the area of interest in order to search for
ambient atomic gas in the local neighbourhood of the SNR. The IAR
30-m single dish telescope at Villa Elisa, Argentina was used for
these observations. The receiver was a helium-cooled HEMT
amplifier with a 1008-channel autocorrelator at the backend, which
has been used for similar purposes by Combi et al. (1998a, b).
System parameters and additional details of the observational
technique can be found in these papers.

The HI line was observed in hybrid total power mode and the sky
was sampled on a $0.5^{\circ}$ rectangular grid. Each grid
position was observed for 36~s with a velocity resolution of
\mbox{$\sim 1$ km s$^{-1}$} and a coverage of \mbox{$\pm 450$ km
s$^{-1}$}. A set of HI brightness temperature maps (\mbox{$\Delta
T_{\mbox{\scriptsize rms}} \sim 0.2$ K}) were built for the
velocity interval (\mbox{$-30$ km s$^{-1}$}, \mbox{$+50$ km
s$^{-1}$}), which is appropriate for this study. We adopted the
following criteria to identify standard clouds in this velocity
range: the maps should show a closed, simply connected structure
(i.e. no shells or filaments were considered), with a clear
maximum in brightness temperature, and an extension in velocity of
at least \mbox{6 km s$^{-1}$}. There are six structures that
satisfy this criterion in the region of interest. The two clouds,
C1 centred at {$(l,\;b)\sim(10^{\circ},\;-12^{\circ})$} and C2
centred at {$(l,\;b)\sim(5.5^{\circ},\;-8^{\circ})$}, are the
stronger ones, with brightness temperatures of {$\sim 45$ K, which
are detected at a confidence level of $225\sigma$}. Figure~4 shows
a series of HI brightness temperature maps where these clouds can
be clearly appreciated. In Figure~5 we present the integrated
column density map for the relevant velocity interval with the 3EG
sources superposed. Two of the sources (3EG~J1834-2803 and
3EG~J1850-2652) are positionally coincident with the clouds C1 and
C2.  Standard galactic rotation models (Fich et al. 1989) indicate
that the remaining clouds, not shown in Figure~4, are located at
kinematic distances that are far too distant to be physically
related to the continuum radio source, which due to its large size
is constrained to be nearby. It is also interesting to notice that
there is a local hole in the HI distribution at the approximate
position of the center of the SNR.
\begin{figure}
\caption{Spectral index distribution computed between 408~MHz and
2.326~GHz for the region shown in Figs. 1 and 2. The EGRET
probability contours are superposed.} \label{fig.3}
\end{figure}

\section{Main results}

The newly-identified extended shell-like radio source in the
Capricornus region is centred at
$(l,b)\approx(+6.5^{\circ},-12.0^{\circ})$. It has an integrated
flux density of \mbox{$\sim180$~Jy} at 408~MHz, once the
contribution from background point sources has been removed. This
extragalactic contribution is dominated by the sources
PKS~1839-34, PKS~1848-338, and PKS~1853-300. Flux errors are
difficult to estimate due to uncertainties related to the
filtering technique, in such a way that the quoted flux should be
considered only as a rough value.

The identification of the extended source as a new SNR is not only
supported by its large size and shell-like morphology, but also by
the spectral indices of the radio emission (see Fig. 3) which are
clearly nonthermal with an average value of
\mbox{$\alpha\approx-0.68\pm0.15$}, somehow steeper than those
observed in typical of shell-type remnants (Green 2000). Notice,
however, that the quoted errors are purely formal and, as it was
previously mentioned, the real ones could be significantly larger
(perhaps around 0.3). The measured properties of this new
Capricornus SNR are summarized in Table~1.
\begin{table}
\caption[]{Measured properties of the new SNR in Capricornus}
\begin{flushleft}
\begin{tabular}{l c}
\noalign{\smallskip} \hline
Property    &  Value \cr \hline Galactic coordinates (center) &
($+6.5^{\circ}$, $-12.0^{\circ}$) \cr Angular size (deg
$\times$deg) & $8.0 \times 8.0$ \cr Flux density (408 MHz) &
$\sim180$ Jy \cr Flux density (2326 MHz) & $\sim55$ Jy \cr Average
spectral index  & $-0.68 \pm 0.15^1$ \cr \hline
\multicolumn{2}{l}{$^1$Formal error.}\cr
\end{tabular}
\end{flushleft}
\end{table}

The distance to the remnant is uncertain, although its large size
and high latitude suggest that it is nearby. A rough distance
estimate of \mbox{$d\sim470$ pc} can be obtained using the
\mbox{$\Sigma (D)$}-dependence for low-brightness shell-like
remnants proposed by Allakhverdigev et al. (1986), although this
method alone is too dubious as to provide a reliable result. An
independent upper limit of 700~pc is provided by the relatively
high value of $|b|$, if the source is assumed to be located within
a galactic disk with a half-height of 150~pc. In addition,
galactic rotation curve models (e.g. Fich et al. 1989) impose an
upper bound of \mbox{$\sim 500$ pc} on the distance, assuming that
the HI clouds C1 and C2 are physically related to the remnant.

A radius of \mbox{$\sim30$ pc} is obtained for the SNR if a
distance, not unreasonable, of 470~pc is adopted, indicating that
it is probably at the end of its adiabatic phase. Using standard
Sedov solutions for an initial energy release of \mbox{$\sim
10^{51}$ erg}, we derive an age of \mbox{$\sim 16,500$ yr},
assuming an average density of \mbox{$n\sim0.05$ cm$^{-3}$} for
the intercloud medium at high galactic latitudes.

In Figure~4 we show the HI clouds C1 and C2 using an ensemble of
channel maps for the velocity interval $-3$ to
\mbox{$+7$~km~s$^{-1}$}, but one of the clouds is visible even
beyond these velocity limits. C1 can be traced from $-8$ to
\mbox{$+2$ km s$^{-1}$} whereas C2 extends from $-1$ to
\mbox{$+5$ km s$^{-1}$}. The cloud column densities were
evaluated by integrating over these velocity intervals.

The integrated HI column density maps were used to obtain
estimates for the total masses of the individual clouds. Using the
adopted distance of 470~pc we estimate masses of \mbox{$\sim 1400$
$M_{\odot}$} and \mbox{$\sim 1200$ $M_{\odot}$} for C1 and C2,
respectively. The aggregate systematic error in these computed
masses may be as large at 50~\%. Dominant sources of error are
uncertainties in the adopted distance, imperfect background
subtraction, and possible existence of other (yet undetected)
molecular species.
\begin{figure}
\caption{HI brightness temperature channel maps (contour labels in
K) obtained for the velocity range $-3.0$ km s$^{-1}$ to $+7.0$ km
s$^{-1}$. The supernova remnant boundary and the EGRET probability
contours are superposed.} \label{fig.4}
\end{figure}

\section{Discussion}

The three $\gamma$-ray sources superposed with the SNR are
classified as being possibly extended in the 3EG catalog, and none
of them show variability. Variability is not expected if the
$\gamma$-ray emission is related to the radio remnant.

The sources 3EG~J1850-2652 and 3EG~J1834-2803 are spatially
coincident with the HI clouds C1 and C2, respectively.  We suggest
that the collective spatial coincidence of the $\gamma$-ray
sources, HI clouds and the radio remnant provides evidence for the
production of $\gamma$-rays by $\pi^0$-decay, where the pions
result from proton-proton collisions at the SNR/ambient cloud
interfaces.

In order to make quantitative estimates of the probability of
chance association of the $\gamma$-ray sources in clusters, we
have adapted the numerical code developed by Romero et al. (1999)
and used by Romero, Benaglia, \& Torres (1999) to study the
positional association of unidentified EGRET sources with various
populations of galactic objects. The code calculates angular
distances between different kinds of celestial objects contained
in selected catalogues, and establishes the level of positional
correlation between them. Monte-Carlo simulations using large
numbers of synthetic populations are then performed in order to
determine the probabilities of pure chance associations. When
generating synthetic populations of $\gamma$-ray sources the
distribution in galactic latitude is constrained to be the same as
that for the actual 3EG sources. This is necessary in order to
obtain reliable results since the distribution of the 3EG sources
is non-isotropic, with a strong concentration towards the galactic
plane. The reader is referred to the paper by Romero, Benaglia, \&
Torres (1999) for further details of the simulation code.

Using this code we determined that there are 54 unidentified 3EG
sources (out of 116) that cluster within $6^\circ$ of each other
at high latitudes (\mbox{$|b|>5^{\circ}$}). When 1500 random
populations of sources are generated (a larger number does not
significantly change the results), the average number of sources
in $6^\circ$ clusters is found to be \mbox{$37.6 \pm 4.7$}. This
translates to a probability as low as \mbox{$1.9\times10^{-4}$}
that all high-latitude $6^\circ$ clusters are chance events. For
\mbox{$|b|<5^{\circ}$} there are 47 sources out of 54 in $6^\circ$
clusters, whereas the simulations predict \mbox{$34.6\pm3.4$}
chance associations.  The probability that all of the observed
low-latitude clusters are chance events is
\mbox{$1.5\times10^{-4}$}, which is similar to the value obtained
for \mbox{$|b|>5^{\circ}$}. These simulations show that several
apparent clusters of unidentified $\gamma$-ray sources are
probably groups of physically related sources.

We shall now estimate the probability of finding a SNR by chance
in the direction of a southern cluster of $\gamma$-ray sources. In
order to do this, we need to know the chance probability of
finding a shell-type, relatively large SNR overlapping with a
circle of the cluster size in an arbitrary direction at
intermediate (say $5^{\circ}<|b|<20^{\circ}$) galactic latitudes.
There is just one of such SNRs in Green's (2000) catalog. Of
course, many large and yet undetected remnants of low surface
brightness should exist under the diffuse non-thermal galactic
emission. A large-scale study with a systematic subtraction of the
galactic background component of the southern radio sky at 2.326
GHz by Jonas (1999) has recently shown that the number of these
uncataloged SNRs in the relevant latitudes is about 17. Since we
have used data from the same Rhodes/HartRAO SKYMAP survey, we
shall adopt this figure as a convenient estimate of the total
number of remnants expected in the region of interest here. Then,
the Poisson probability of finding one SNR overlapping with a
cluster of 3 or more unidentified 3EG sources is $\sim1.4\times
10^{-2}$.

The probability of chance association is further constrained if we
consider that there are two clouds apparently coincident with two
of the three $\gamma$-ray sources in the cluster for which the new
SNR was found. The probability $p(n)$ that the line of sight along
a distance $r$ will intersect with $n$ clouds is (Spitzer 1998):

\begin{equation}
p(n)=\frac{(kr)^n e^{-kr}}{n!},
\end{equation}
where $k$ is the average number of clouds per kpc\footnote{This is
actually an oversimplification because it ignores the fact that
clouds form a power-law distribution of properties. The results,
however, are not significantly changed in the present case, where
massive clouds are considered.}. In the direction of the galactic
plane \mbox{$k\sim6.2$} for standard clouds (Spitzer 1998). We
found no published estimates of $k$ for relatively high latitudes,
so we adopt the galactic plane value in order to obtain an upper
limit to the probability. Using this equation and a distance range
equal to the estimated diameter of the SNR (60 pc), we obtain that
the chance probability of finding one or more clouds in each of
two different lines of sight is 0.097 (about 10 \%). Consequently,
the probability that the observed configuration of SNR,
$\gamma$-ray sources and HI clouds be the sole effect of chance
superposition is about $1.3\times10^{-3}$. This probability,
although not overwhelming, suggests that the hypothesis of a
multiple $\gamma$-ray source that could be causally related to the
SNR deserves serious study in the light of its important potential
implications.

There are mechanisms other than $p-p$ induced $\pi^0$ decay that
may contribute to the $\gamma$-ray emission from SNRs. Inverse
Compton (IC) scattering of relativistic electrons can boost
ambient low-energy photon fields (such as the cosmic microwave
background radiation, the diffuse galactic infrared/optical
radiation, and the remnant's own photon fields) to $\gamma$-ray
energies (e.g. Pohl 1996, Mastichiadis 1996). Bremsstrahlung
emission from electrons in the shocked gas can also yield
significant $\gamma$-ray luminosities (e.g. Blandford \& Cowie
1982). The relative contribution of each of these processes to the
overall $\gamma$-ray flux from the remnant depends on factors such
as the age of the remnant, the injected particle spectra, the
ambient gas density, and the ratio of electrons to protons at high
energies (e.g. Sturner et al. 1997, Gaisser et al. 1998).

In large and old remnants, such as the Capricornus remnant
presented here, the IC boosting of diffuse background photons
should dominate that of local radiation fields (Sturner et al.
1997). At high energies electrons cool more efficiently than
protons by synchrotron and IC losses, therefore the
electron-to-proton ratio in the source is expected to be
\mbox{$\ll 1$}, i.e. it would be similar to what is observed in
the CRs near the Earth (see Aharonian et al. 1994, and especially
Biermann \& Strom 1993 for additional arguments). IC radiation is
therefore not expected to be dominant. Two factors indicate that
the EGRET sources considered in this study are not predominantly
due to IC scattering: (a) the detectable $\gamma$-ray emission is
located towards regions where the observed radio emission (and
hence the energy flux of relativistic electrons) is not
particularly strong, and (b) two of the $\gamma$-ray sources occur
towards HI clouds.
\begin{figure}
\caption{A map of integrated column density of HI (velocity
interval from $-3.0$ km s$^{-1}$ to $+5.0$ km s$^{-1}$) for the
region of interest. Label units are $10^{19}$ atoms cm$^{-2}$.
EGRET probability contours are superposed.} \label{fig.5}
\end{figure}
If the HI clouds are sufficiently dense and the relativistic
electron flux is high enough, then the observed $\gamma$-ray flux
could be produced by bremsstrahlung interactions between the
electrons and the ionized atoms in the compressed post-shock
regions. This mechanism can, however, be discounted for our
sources because both the cloud densities (\mbox{$n\sim10$
cm$^{-3}$}, inferred from the HI data) and the electron flux
(estimated from the radio synchrotron emission) are far too low.
The electron flux necessary to produce the observed $\gamma$-ray
luminosity by bremsstrahlung mechanism would result in a
synchrotron radio flux of about 100 Jy at 1 GHz for each cloud.
This radio flux is two orders of magnitude greater than the value
determined from the radio data for the regions occupied by the
clouds in the remnant, and more than the flux density of the
entire source at that frequency. We conclude, then, that $\pi^0$-
decay is the dominant mechanism for $\gamma$-ray production in the
new remnant. When an expanding SNR collides with a cloud, a
fraction of the nuclear CR component is transported into the cloud
by convection. This penetration produces a significant enhancement
of the $\gamma$-ray luminosity because of the $\pi^0$-decays
resulting from the $p-p$ interactions. The $\gamma$-ray flux
expected at the Earth is:
\begin{equation}
F(E>100\;{\rm MeV})=\frac{1}{4\pi} q_{\gamma} M_{\rm cloud}\;
d^{-2} m_{\rm H}^{-1},
\end{equation}
where $q_{\gamma}$ is the $\gamma$-ray emissivity per H atom of
the cloud of mass $M_{\rm cloud}$, $d$ is the distance to the
source, and $m_{\rm H}$ is the H-atom mass.

Using the observed $\gamma$-ray fluxes from the C1 and C2
positions, the adopted distance of 470 pc, and our cloud mass
estimates derived from the HI-observations, we find that there is
an enhancement in the CR energy density in the clouds with respect
to the value observed near the Earth (Dermer 1986). The
enhancement factor is calculated to be \mbox{$\sim 4.5$} for cloud
C1 and \mbox{$\sim 13$} for cloud C2, clearly indicating that
protons are being accelerated at the remnant shock front.

Since at high energies the spectrum of $\pi^0$-decay $\gamma$-rays
mimics the spectrum of the parent protons, we can infer from EGRET
observations that at the position of C2 the proton spectrum is
$N(E_p)\propto E^{-2.6\pm0.2}$. This power-law index is similar to
that of the electron component at the same position, as deduced
from the radio brightness spectral index. The $\gamma$-ray
spectrum towards C1 may be harder: $N(E_p)\propto
E^{-2.3\pm0.45}$. The radio emission from this region is too weak
to provide a reliable electron spectral index for comparison. In
both cases the proton spectral index is consistent with the
theoretical value of \mbox{$\Gamma=2.42$} that is expected from
diffusive shock acceleration in shell-like remnants (Biermann \&
Strom 1993; see also Gaisser et al. 1998, who consider a harder
injection spectrum with \mbox{$\Gamma=2.3$}). Leakage from the
galactic disk results in the steeper $E^{-2.75}$-spectrum observed
near the Earth (Biermann \& Strom 1993). We estimate that the
proton high-energy cutoff, considering the age of the SNR, should
be about 80~TeV (see Biermann \& Strom 1993 for the formulae).

\section{Conclusions}

We have discovered a new southern SNR in the Capricornus region.
It is a large and faint shell-type nonthermal radio source which
has till now been obscured by confusion with the diffuse galactic
radiation. There are three $\gamma$-ray sources in the 3EG catalog
superimposed on this remnant. Two of these high-energy sources
coincide with the positions of HI clouds detected at 21 cm in
narrow velocity intervals. We have estimated the masses of the
clouds and calculated the CR flux enhancement at the SNR, showing
convincing evidence of proton acceleration at the shock front, as
predicted by the theory. Interestingly, it is possible that the
proton spectrum slope is not the same at all points in the
remnant. The electron and proton components of the CR flux seem to
display similar spectral slopes towards the single position where
both spectra can be determined. Although the possibility of a pure
chance association cannot be completely ruled out, it is clear
that this southern cluster of $\gamma$-ray sources deserves more
attention as a potential natural laboratory for CR studies. In
particular, future X-ray and TeV observations of the new
Capricornus SNR are required to provide the information necessary
to determine the upper cutoffs in the spectrum of both the
leptonic and the hadronic CR components within the source.

\begin{acknowledgements}

We are very grateful to Dr. D.F. Torres for his kind help with the
numerical simulations and to an anonymous referee for useful
comments. This research has been supported by the Argentine
agencies CONICET (under grant PIP N$^o$ 0430/98) and ANPCT (PICT
98 No. 03-04881), as well as by Fundaci\'on Antorchas (funds
granted to J.A.C. and G.E.R.). J.L.J. acknowledges financial
support from the Joint Research Committee of Rhodes University.

\end{acknowledgements}

{}

\end{document}